\documentclass[notitlepage,aps,prx,twocolumn,twoside,superscriptaddress,showkeys,nofootinbib]{revtex4-1}
\usepackage{hyperref}
\usepackage{graphicx}
\usepackage{amsmath,amsfonts,amssymb,amsthm,amsbsy,mathtools}
\usepackage{bm}
\usepackage{dsfont}
\usepackage{bbold}
\usepackage{epic}
\usepackage{eepic}
\usepackage{mathrsfs}
\usepackage{lmodern}
\usepackage[T1]{fontenc}
\usepackage[utf8]{inputenc}
\usepackage{color}
\usepackage[english]{babel}

\newcommand{\ket}[1]{| {#1} \rangle} 
\newcommand{\braket}[2]{\langle {#1} \vphantom{#2} | {#2} \vphantom{#1} \rangle} 
\newcommand{\proj}[1]{|#1\rangle\langle#1|}

\DeclareMathOperator*{\tr}{\mathrm{Tr}}

\begin{document}

\title{Wigner's friend as a rational agent}
\author{Veronika Baumann$^{1,2}$ and {\v C}aslav Brukner}
\affiliation{Vienna Center for Quantum Science and Technology (VCQ), Faculty of Physics, University of Vienna, Boltzmanngasse 5, A-1090 Vienna, Austria}
\affiliation{Institute of Quantum Optics and Quantum Information (IQOQI), Austrian Academy of Sciences, Boltzmanngasse 3, A-1090 Vienna, Austria}
\date{\today}

\begin{abstract}

In a joint paper Jeff Bub and Itamar Pitowski argued that the quantum state represents  {\em ``the credence function of a rational agent [...] who is updating probabilities on the basis of events that occur''}. In the famous thought experiment designed by Wigner, Wigner's friend performs a measurement in an isolated laboratory which in turn is measured by Wigner. Here we consider Wigner's friend as a rational agent and ask what her ``credence function'' is. We find experimental situations in which the friend can convince herself that updating the probabilities on the basis of events that happen solely inside her laboratory is not rational and that conditioning needs to be extended to the information that is available outside of her laboratory. 
Since the latter can be transmitted into her laboratory, we conclude that the friend is entitled to employ Wigner's perspective on quantum theory when making predictions about the measurements performed on the entire laboratory, in addition to her own perspective, when making predictions about the measurements performed inside the laboratory. 
\end{abstract}

\maketitle
\setcounter{page}{1}
Wigner's-friend-type experiments comprise observations of observers who themselves have performed measurements on other systems. In the standard version of the thought experiment~\cite{wigner1963problem} there is an observer called Wigner's friend measuring a quantum system in a laboratory, and a so-called superobserver Wigner,  to whom the laboratory including the friend constitutes one joint quantum system. From the point of view of the friend, the measurement produces a definite outcome, while for Wigner the friend and the system undergo a unitary evolution. The discussion surrounding the Wigner's-friend gedankenexperiment revolves around the questions: Has the state of the physical system collapsed in the experiment? Has it collapsed only for the friend or also for Wigner? 

Recently, extended versions of Wigner's-friend scenarios have been devised in order to show the mutual incompatibility of a certain set of (seemingly) plausible assumptions. In Ref.~\cite{brukner2015quantum,brukner2018no} a combination of two spatially distant Wigner's-friend setups is used as an argument against the idea of ``observer-independent facts'' via the violation of a Bell-inequality. In Ref.~\cite{frauchiger2018quantum} two standard Wigner's-friend setups are combined such that one arrives at a contradiction when combining the supposed knowledge and reasoning of all the agents involved. Both arguments are closely related to the quantum measurement problem and attempted resolutions involve formal as well as interpretational aspects~\cite{brukner2015quantum,baumann2017formalisms}. At the core of the formal aspects is the standard measurement-update rule. Without further ontological commitment regarding a ``collapse of the wave function'' the latter can be considered to be the rational-update-belief rule for agents acting in laboratories. The updated state is the one that gives the correct probabilities for subsequent measurements and would, hence, be used by a rational agent to make her predictions.

Here we regard Wigner's friend as a rational agent and not only as a (complex) quantum system, which is normally considered in the literature. Acting as such an agent, she aims to achieve a rational prediction regarding the outcome of any future measurement. 
We find a situation in which the friend can verify that her prior prediction about future measurement outcomes --  which is based on applying the standard state-update rule on the basis of the events that happen inside her laboratory -- is incorrect. In this way, the friend can convince herself, that her prediction cannot be conditioned solely on the outcomes registered inside her laboratory, but needs to take into account the information that is available outside of the laboratory as well. 
The latter can be transmitted to her by Wigner or an automatic machine from outside her laboratory. 
Equipped with this additional information the friend is entitled to adopt both Wigner's and her own perspective on the use of quantum theory.

We first consider a simple version of our Wigner's-friend experiment, moving later to a more general case. Let a source emit a qubit~$S$ (spin-1/2 particle) in state $|\phi\rangle_S= 1/\sqrt{2} (|\uparrow\rangle_S + |\downarrow\rangle_S)$, where $|\uparrow\rangle_S $ is identified with ``spin up'' and $|\downarrow\rangle_S$ with ``spin down'' along the $z$-direction. The friend $F$ measures the particle in the $\sigma_z$-basis (Pauli $z$-spin operator) with the projectors $M_F$: $\{ |\uparrow\rangle \langle \uparrow|_S, |\downarrow\rangle \langle \downarrow|_S \}$. To Wigner $W$, who is outside and describes the friend's laboratory as a quantum system, the measurement of the qubit by the friend constitutes a unitary evolution entangling certain degrees of freedom of the friend -- her memory that is initially in state $|0\rangle$ (i.e. the register in state ``no measurement'') -- with the qubit.
Suppose that this results in the overall state 
\begin{equation}
\label{eq:unitary}
|\phi\rangle_S |0\rangle_F \mapsto \frac{1}{\sqrt{2}} (|\uparrow\rangle_S |U\rangle_F + |\downarrow\rangle_S |D\rangle_F)=: |\Phi^{+}\rangle_{SF},
\end{equation}
where $|Z\rangle_F$ with $Z\in\{U, D \}$ is the state of the friend having registered outcome $z\in\{\text{``up''}, \text{``down''}\}$ respectively. Wigner can perform a highly degenerate projective measurement $M_W$: $\{ |\Phi^{+}\rangle \langle \Phi^{+}|_{FS}, \mathds{1} -|\Phi^{+}\rangle \langle \Phi^{+}|_{FS} \}$ to verify his state assignment. The respective results of the measurement will occur with probabilities $p(+)=1$ and $p(-)=0$. We note that the specific relative phase between the two amplitudes in Eq. (\ref{eq:unitary}) (there chosen to be zero) is determined by the interaction Hamiltonian between the friend and the system, which models the measurement and is assumed to be known to and in control of Wigner. 

\onecolumngrid
\begin{center}
\begin{figure}[h] 
\includegraphics[width=18.5cm]{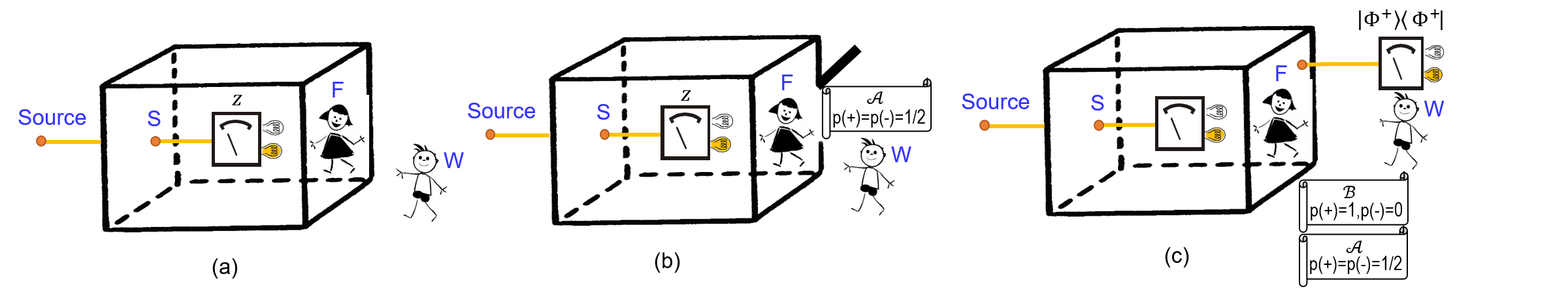}
\caption{Three steps of the protocol: (a) Wigner's friend (F) performs a measurement $M_F$ on the system S in the laboratory and registers an outcome. (b) The friend makes a prediction about the outcome of a future measurement $M_W$ on the laboratory (including the system and her memory), for example, by writing it down on a piece of paper ${\cal A}$. She communicates her prediction to Wigner (W). (c) Wigner performs measurement $M_W$. The three steps (a-c) are repeated until sufficient statistics for measurement $M_W$ is collected. The statistics is noted, for example, on another piece of paper ${\cal B}$. Afterwards, the ``lab is opened'' and the friend can compare the two lists ${\cal A}$ and ${\cal B}$.  An alternative protocol is given in the main text.}
\end{figure}
\vspace{-0.75em}
\end{center}    
\twocolumngrid


We next introduce a protocol through which Wigner's friend will realise that her predictions will be incorrect, if she bases them on the state-update rule conditioned only on the outcomes observed inside her laboratory. The protocol consists of three steps (see Fig. 1): (a) The friend performs measurement $M_F$ and registers an outcome; (b) The friend makes a prediction about the outcome of measurement $M_W$ and communicates it to Wigner; (c) Wigner performs measurement $M_W$. Besides the outcome of $M_F$ measurement no further information is given to the friend.\\

Step (a): Following the laboratory practice and textbook quantum mechanics, when the friend observers outcome $z$, she applies the measurement-update rule, and consequently predicts the probabilities: $p(+|z)=|\langle z| \langle Z| \Phi^{+}\rangle_{SF}|^2= \frac{1}{2} = p(-|z)$, where $|z\rangle_S | Z\rangle_F $ can be either $|\uparrow\rangle_S |U\rangle_F$ or $|\downarrow\rangle_S |D\rangle_F$ depending on which outcome the friend observes. It is important to note that the friend's prediction of the result of $M_W$ does not reveal the result of $M_F$, because for either outcome $z$ the predicted conditional probability is the same, i.e. $p(+)=p(-)= \frac{1}{2}$. This means that the friend's prediction could be sent out of the laboratory and communicated to Wigner without changing the superposition state $|\Phi^+\rangle_{SF}$.\\ 

Step (b): The friend opens the laboratory in a manner that allows a record of her prediction (e.g., a specific message $\cal{A}$ written on a piece of paper) to be passed outside to Wigner, keeping all other degrees of freedom fully isolated and in this way preserving the coherence of the state. The piece of paper waits outside of the laboratory to be evaluated later in the protocol. The state that Wigner assigns to the friend's laboratory at the present stage is $ |\Phi^{+}\rangle_{SF} |p(+)\!=\!p(-)\!=\!1/2\rangle_{\cal A}$, where the second ket refers to the message. \\

Step (c): After the friend communicates her prediction to the outside, Wigner (or an automatic machine) performs measurement $M_W$ and records the result, for example on a separate piece of paper $\cal{B}$. At this stage of the experiment, Wigner assigns to the friend's laboratory state and the two messages the state $ |\Phi^{+}_d\rangle_{SF} |p(+)\!=\!p(-)\!=\!1/2\rangle_{\cal A} |p(+)\!=\!1,p(-)\!=\!0\rangle_{\cal B}$ with two last kets representing the two lists. \\

We note that the measured state $|\Phi^+\rangle_{SF}$ is an eigenstate of the measured operator. This implies that outcome ``+'' will occur with unit probability, and furthermore that the measurement does not change the state of the laboratory. Hence the three steps (a-c) of the protocol can be repeated again and again without changing the initial superposition state until a sufficient statistics in measurement $M_W$ is collected. As a result, the two pieces of paper, one with the friend's prediction $ {\cal A} = \{ p(+)=\frac{1}{2}, p(-)=\frac{1}{2} \}$, and one with the actually observed relative number of counts for the two outcomes ${\cal B} = \{p(+)=1, p(-)=0 \}$, display a statistically significant difference. In the very last step of the protocol ''the laboratory is opened'' (which is equivalent to Wigner performing a measurement in the basis $\{ |z\rangle\langle z|_S \otimes |Z'\rangle \langle Z'|_F \}$), such that the joint states for the systems and the friend's memory are reduced to the form $|z\rangle_S |Z\rangle_F$. Irrespective of the specific state to which the friend is reduced, she can now compare the two messages and convince herself that her prior prediction deviates from the actually observed statistics. 

In an alternative protocol the friend resides in the laboratory for the duration of the whole protocol. Steps (a) and (c) are retained, but instead of step (b), the friend either receives the result of measurement $M_W$ from Wigner after each measurement run or the entire list ${\cal B}$ at the end of the protocol. In either case the friend can compare the two lists inside the laboratory and arrive at the same conclusion as in the previous protocol. One may object that in the present case the friend's conclusion relies on Wigner being trustful and reliable when providing her with the measurement result of $M_W$. As mentioned above, this is irrelevant as Wigner can be replaced in the protocol by an automatic machine that the friend could have pre-programmed herself.

The discrepancy between the friend's prediction and the actual statistics can be made as high as possible in a limit. Consider that the source emits a higher dimensional quantum system in state $|\phi_d\rangle_S=\frac{1}{\sqrt{d}}\sum_{j=1}^{d} |j\rangle_S$ and the friend measures in the respective basis, $M_F: \{|j\rangle\langle j| _S \}$ with $j=1\dots d$. Wigner describes the measurement of the friend as a unitary process that results in state
\begin{equation}
|\phi_d\rangle_S |0\rangle_F \mapsto \frac{1}{\sqrt{d}} \sum_{j=1}^{d} |j\rangle_S |\alpha_j\rangle_F =: |\Phi^{+}_d\rangle_{SF},
\label{eq:unitary1}
\end{equation}
where $|\alpha_j\rangle_F$ is the state of the friend's memory after having observed outcome $j$. For $M_W$ we choose: $\{ |\Phi^{+}_d\rangle \langle \Phi^{+}_d|_{SF}, \mathds{1}-|\Phi^{+}_d\rangle \langle \Phi^{+}_d|_{SF} \}$. Wigner records the ``+'' result with unit probability, i.e. $p(+)=1$ and $p(-)= 0$. Using the measurement-update rule the friend, however, predicts 
\begin{equation}
\begin{aligned}
p(+|j)&=|\langle j|\langle \alpha_j|\Phi^{+}_d\rangle_{SF}|^2=\frac{1}{d} \quad \underset{d\rightarrow \infty}{\longrightarrow}\quad0\\
p(-|j)& = 1 - \frac{1}{d} \underset{d\rightarrow \infty}{\longrightarrow}\quad1 \;,
\end{aligned}
\label{eq:pF}
\end{equation} 
independently of the actual outcome she registers in measurement $M_F$. 

The three steps (a-c) of the protocol can be repeatedly applied. In that way, the friend can convince herself, without relying on the knowledge other agents may have, that she made an incorrect prediction of future statistics using the state-update rule conditioned solely on the observations made within her laboratory. As the dimension of the measured system and the number of outcomes increases, the discrepancy between her prediction and the actual statistics becomes maximal, giving rise to an ``all-versus-nothing'' argument. 

We now discuss possible responses of different interpretations and modifications of quantum theory to the experimental situation described above.

(1) {\it Theories that deny universal validity of quantum mechanics.} The theories postulate a modification of standard quantum mechanics, such as spontaneous~\cite{ghirardi1986unified} and gravity-induced \cite{penrose2000wavefunction,diosi2014gravity} collapse models, which become significant at the macroscopic scale. These models could exclude superpositions of observers as required in Wigner's-friend-type experiments and hence the protocol cannot be run. In our view this is the most radical position as it would give rise to new physics. 

(2) {\it Many-worlds interpretation}~\cite{wheeler1957assessment}. The incorrect prediction of the friend is due to her lack of knowledge of the total state of the laboratory including the friend's memory (i.e. ``the wave function of the many-worlds''). If the friend would know this state, she  could make a prediction that is in accordance with actually observed statistics. 

(3) {\it Quantum Bayesian}~\cite{fuchs2010qbism}, {\it neo-Copenhagen}~\cite{brukner2015quantum}, {\it and relational interpretation}~\cite{rovelli1996relational}. The quantum state is always given relative to an observer and represents her/his knowledge or degree of belief. It is natural for an agent to update his or her  degree of belief in face of new information. As a consequence of newly acquired knowledge in the protocol the friend would update her degree of belief and assign a new state to the laboratory that is in agreement with the observations. 

Note that in both interpretations (2) and (3) the friend {\it can learn} the state of the laboratory as described by Wigner. Wigner can simply communicate either the initial state $|\phi\rangle_S |0\rangle_F$  of the system together with  the Hamiltonian of the laboratory, or the final state $|\Phi^{+}_d\rangle_{SF}$ of  the laboratory to the friend before the protocol begins. The important feature is that this can be done without destroying the coherence of the quantum state of the laboratory. In this way, the friend can learn the overall state of her laboratory from the perspective of an outside observer and include it in making her prediction. 
As a consequence, the friend may operate with a pair of states $\{ |z\rangle_S, |\Phi^{+}_d\rangle_{SF} \}$. The first component is used for predictions of measurements made on the system alone and is conditional on the registered outcome in the laboratory; the second component is used for predictions of measurements performed on the entire laboratory (= ``system + friend''). When an outcome of a measurement on the system is registered in the laboratory, the friend applies the state-update rule on the first component, without affecting the second component of the pair of states. 

Alternatively, the friend can use only the quantum state Wigner assigns to the laboratory, but apply a modification of the Born rule as introduced in Ref.~\cite{baumann2017formalisms}. The new rule enables to evaluate conditional probabilities in a sequence of measurements. In the case when the sequence of measurements is performed on the same system, it restores the predictions of the standard state-update rule. However, the rule enables making predictions in Wigner's-friend scenarios as well, where the first measurement is performed on the system by Wigner's friend, and the subsequent measurement is performed on the ``system + friend'' by Wigner. In this case, the rule recovers the prediction as given by Eq.~\eqref{eq:pF}. This modified Born rule gains an operational meaning in the present setup where Wigner's friend has access to outcomes of both measurements and can evaluate the computed conditional probabilities (Note that Wigner does not have this status, since he has no access to the outcome of the friend's measurement.). In terms of Bub and Pitowski, both the ``outside'' and ``inside'' perspective contribute to ``the credence function'' of Wigner's friend. 

\begin{appendix}
\section{Modified Born rule~\cite{baumann2017formalisms}}

We apply the modified Born rule of Ref.~\cite{baumann2017formalisms} to the present situation. The rule defines the conditional probabilities in a sequence of measurements. According to it, every measurement is described as a unitary evolution analogous to Eq.~\eqref{eq:unitary}. Consider a sequence of two measurements, the first one described by unitary $U$ and the subsequent one by unitary $V$. The sequence of unitaries correlate the results of the two measurements with two memory registers 1 and 2. The overall state of the system and the two registers evolve as 
\begin{eqnarray}
\label{eq:Phi_tot}
\ket{\phi}_S \ket{0}_1 \ket{0}_2 & \xrightarrow{U} &  \sum_{j} \braket{j}{\phi} \ket{j}_S \ket{\alpha_j}_1 \ket{0}_2 \\
 & \xrightarrow{V} & \sum_{jk} \braket{j}{\phi} \langle \tilde\beta_{jk}|j, \alpha_j\rangle \ket{\tilde \beta_{jk}}_{S1} \ket{\beta_{k}}_2 = \ket{\Phi_{tot}}. \nonumber
\end{eqnarray}
The conditional probability  for observing outcome $\bar{k}$ given that outcome $\bar{j}$ has been observed in the previous measurement is given by 
\begin{equation}
p(\bar{k}|\bar{j}) = \frac{p(\bar{k},\bar{j})}{\sum_{\bar{k}} p(\bar{k},\bar{j})} . 
\end{equation}
Here $p(\bar{k},\bar{j})$ is the joint probability for observing the two outcomes and is given by the projection of the overall state $\ket{\Phi_{tot}}$ on the states $\ket{\alpha_{\bar j}}_{1}$ and $\ket{\beta_{\bar k}}_2$ of the two registers:
\begin{eqnarray}
p(\bar{j},\bar{k})= \tr[(\mathds{1}_S\otimes \proj{\alpha_{\bar j}}_1 \otimes \proj{\beta_{\bar k}}_2) \proj{\Phi_{tot}}] \nonumber \\
\label{eq:EBorn}
\end{eqnarray}
In the case when the sequence of the measurements are performed on a single system, the rule restores the prediction of the standard state-update rule, i.e. $p(\bar{k},\bar{j})=|\langle \bar{k}|\bar{j}\rangle|^2$. This corresponds to the state
\begin{equation}
\ket{\Phi_{tot}} = \sum_{jk} \braket{j}{\phi} \braket{k}{j} \ket{k}_S \ket{\alpha_j}_1 \ket{\beta_{k}}_2  .
\end{equation}

For Wigner's-friend scenarios, however, the friend would predict conditional probabilities for Wigner's result $+$ given her result $j$ that are in accordance with Wigner's observations Eq.~\eqref{eq:pF}. In this case the overall state is
\begin{equation}
\ket{\Phi_{tot}}= \sum_{j=1}^{d} \braket{j}{\phi} \ket{j}_S \ket{\alpha_j}_{1} \ket{+}_2
\end{equation}
where $\ket{+}_2$ is the state of the memory register corresponding to observing the result of $\proj{\Phi^+_d}_{S1}$, and the joint probability is given by
\begin{equation}
p(+,j) = \tr[(\mathds{1}_S \otimes \proj{\alpha_j}_1 \otimes \proj{+}_2) \proj{\Phi_{tot}}].
\end{equation}
This means that full unitary quantum theory together with the adapted Born rule allows for consistent predictions in Wigner's-friend-type scenarios, while giving the same predictions as the standard Born and measurement-update rules for standard observations.
\end{appendix}


{\it Acknowledgments} We thank Borivoje Dakic, Flavio del Santo, Renato Rennner and Ruediger Schack for useful discussion. We also thank Jeff Bub for the discussions and for the idea of the alternative protocol. We acknowledge the support of the Austrian Science Fund (FWF) through the Doctoral Programme CoQuS (Project no. W1210-N25) and the projects I-2526-N27 and I-2906. This work was funded by a grant from the Foundational Questions Institute (FQXi) Fund. The publication was made possible through the support of a grant from the John Templeton Foundation. The opinions expressed in this publication are those of the authors and do not necessarily reflect the views of the John Templeton Foundation.

\bibliography{WignerRef}
\end{document}